# CLASSICAL UNIVERSES AND QUANTIZED PARTICLES FROM FIVE-DIMENSIONAL NULL PATHS

Paul S. Wesson

Department of Physics and Astronomy, University of Waterloo, Waterloo, Ontario N2L 3G1, Canada

Correspondence: Mail to (1) above. Email: Space-Time-Matter Consortium, http://astro.uwaterloo.ca/~wesson (for publication); pswastro@hotmail.com (editorial, not for publication).

Abstract

All objects in 4D spacetime may in principle travel on null paths in a 5D manifold. We use this, together with a change in the extra coordinate and the signature of the metric, to construct a simple model of a classical universe and a quantized particle. There are notable implications for the Weak Equivalence Principle and the cosmological constant, and for astrophysics.

1. Introduction

Einstein's equations do not restrict the dimensionality of the world, and it has become commonplace to extend spacetime as a means of unifying gravitational physics with particle physics. In the basic extension, it has been known for several years that a massive particle moving on a timelike path in 4D may in principle be moving on a null path in 5D. It is also possible to change the fifth coordinate and the sign of the fifth part of the metric, resulting in different guises for 4D physics. Here, we wish to use these 5D properties to construct 4D models for a classical universe and a quantized particle. Our account is preliminary in nature; but we will come to the remarkable conclusion that the universe and a particle may be different 4D aspects of the same 5D structure.

The two versions of 5D relativity currently under discussion are membrane theory and induced-matter (or space-time-matter) theory. The former views 4D spacetime as a singular hypersurface in a 5D manifold, thereby explaining the relative strength of particle interactions by confining them to this membrane, whereas weaker gravity can propagate freely into the external bulk. By contrast, induced-matter theory envisages an unconstrained 5D manifold, whose extra effects are identified as 4D matter, which is constrained by the geodesic equation. A review of both approaches may be found in a recent book [1]. Both theories are in agreement with the classical and other tests of relativity; but to work out their consequences in a given situation, it is necessary in general to assume a starting form for the metric. This is frequently taken to be the warp metric for membrane theory and the canonical metric for induced-matter theory. The former involves an exponential factor in the extra coordinate applied to 4D spacetime, while the latter involves a quadratic factor, with scales in both cases set by the effective value of the cosmological constant as it appears in Einstein's equations of general relativity. The field equations for both versions of 5D relativity are taken to be extensions of the 4D Einstein ones. In their simplest form, in terms of the 5D Ricci tensor, these are the relations $R_{AB} = 0$ ($A, B = 0,123,4$ for time, ordinary space and the extra dimension). For induced-matter theory, these are in fact the full equations, since it may be shown that these contain the 4D Einstein equations. In terms of the Einstein tensor and the energy-momentum tensor, these are as usual $G_{\alpha\beta} = 8\pi T_{\alpha\beta}$ ($\alpha, \beta = 0,123$). The properties of matter in this approach depend on the extra metric coefficients and derivatives of the 4D metric coefficients with respect to the extra coordinate, hence the appellation induced-matter (or space-time-matter) theory. This kind of physical embedding is guaranteed in a mathematical sense by Campbell's theorem. This also applies in a modified manner to membrane theory, and it is clear from what has been stated that the two versions of 5D relativity are similar. Indeed, it is now known that they are equivalent in terms of the field equations and the equations of motion [2, 3], even though they differ in motivation and physical form.

In what follows, we will not be much concerned with field equations or matter, but rather with metrics and the equations of motion. Our approach therefore has some generality. However, we will adopt the canonical metric, since it is well suited to our goal. This is to show that classical mechanics and quantum mechanics are not necessarily disparate, but may be viewed as flowing from the same underlying mathematical structure. The main assumption (physical and mathematical) we will make, is that all objects in 4D travel along null paths in 5D. This has been shown viable both for induced-matter theory [4] and membrane theory [5]. In both theories there is also a kind of inertial fifth force which arises if there is relative motion between the 4D and 5D frames [6, 7]. However, our results will be seen to follow mainly from two other properties of the 5D metric in canonical form. Firstly a change in the fifth coordinate represents a kind of duality which may be related to the Planck mass and masses in general [8, 9]. Secondly, a change in the signature of the metric (in which the fifth dimension is taken to be timelike instead of spacelike) allows a switch from the real quantities of classical mechanics to the complex quantities of quantum mechanics. Such so-called two-time metrics have already been shown to be physically acceptable in $N > 4D$ field theory, notably in 5D [10] and 6D [11]. By using these changes in coordinate and signature, in conjunction with the 5D geodesic equation with the null condition, we will be able to construct models for a classical universe and a quantized particle.

Some of the results obtained below are similar to ones derived from a previous analysis of the classical and quantum consequences of 5D relativity [8]. However, we can now take advantage of several new discoveries in the subject. These concern: (a) the gauge dependence of the cosmological 'constant' [12], which may alleviate the problem posed by the mismatch in the size of this parameter as determined from cosmology and particle physics [13]; (b) the astrophysical implications of a variable cosmological 'constant' and other aspects of 5D relativity [14]; (c) the nature of the inflationary phase in 5D cosmology [15]; (d) the topology and structure of $N$D metrics of the canonical sort [16]. We will also make contact with other results, some of which are older and well known. These concern: (i) the 1-body or soliton solutions of 5D gravity [17-20]; (ii) the 5D analogs of the 4D Friedmann-Robertson-Walker solutions of standard cosmology [21, 22]; (iii) other 5D solutions relevant to cosmology and astrophysics [23]; (iv) astrophysical tests of 5D gravity [24,25]; (v) dynamical consequences of the 5D geodesic equation and its relationship to particle mass [26-29]; (vi) the application of Campbell's theorem to the embedding of 4D general relativity in 5D theories of the membrane and induced-matter type [30]. Below, we will employ a number of results which are proved in the papers just listed.

A short summary of the formalism we are using is available for those not familiar with it [31]. Those conversant with the canonical metric and its consequences, or those more interested in physics than mathematics, may skim Section 2. The application of that metric, with a spacelike extra coordinate, to classical mechanics and cosmology occupies Section 3. The corresponding application, with a timelike extra coordinate, to quantum mechanics and particle physics is to be found in Section 4. We review our results and outline future topics for research in Section 5. Throughout, we let upper-case Latin (English) letters run 0-4, and lower-case Greek letters run 0-3. We generally choose units which render the gravitational constant ($G$), the speed of light ($c$) and the quantum of action ($h$) all unity, except in places where we restore them for physical clarity. This work is exploratory, to see how far there may be an overlap between classical and quantized behaviour in 4D from the viewpoint of null paths in 5D. Throughout, we will mean by "quantum" mechanics the older version of that term as applied to a single particle in the manner of Bohr and Sommerfeld, rather than the newer sum-over-paths approach, which is beyond the scope of the present work.

2. Metrics and Their Consequences

Numerous exact solutions are known of the field equations $R_{AB} = 0$ ($A, B = 0 - 4$; see ref. 1). In general, these 15 relations comprise 10 gravity-type equations, 4 electromagnetic-type equations, and 1 wave equation for the scalar field. However, the form of these equations depends on the form assumed for the metric (i.e. on the coordinates or gauge). The canonical metric is so called because it leads to great simplification in the field equations, and in the geodesic equation from which the laws of motion are derived. It does this by setting the potentials of electromagnetic type $(g_{0\alpha})$ to zero, and the magnitude of the potential of scalar type $(g_{44})$ to one. This procedure allows us to concentrate on the 10 Einstein-like relations of the field equations. It uses up the 5 degrees of coordinate freedom available in a 5D metric, but this is still algebraically general as long as we allow the 4D potentials of gravitational type to depend not only on the coordinates of spacetime $(x^\alpha)$, but also on the extra coordinate $(x^4 = l)$. It is this dependency which defines the singular hypersurface in membrane theory, and gives rise to the energy-momentum tensor in space-time-matter theory. (Membrane theory commonly labels the fifth coordinate $x^4 = y$, but this can be confused with the Cartesian coordinate, so here we use $x^0 = t$, $x^{123} = xyz$ or $r\theta\phi$, and $x^4 = l$.) Because of the dependency $g_{\alpha\beta} = g_{\alpha\beta}(x^\gamma, l)$ of the 4D metric tensor, we can also extract a function of $l$ if we wish, and it proves beneficial to extract a quadratic factor $l^2/L^2$, where $L$ is a constant length introduced for the consistency of physical dimensions. This factor also means that the 5D canonical metric is analogous to the 4D Robertson-Walker metric for the Milne universe, which we will see below enables us to draw certain parallels between 5D and 4D. Indeed, the canonical metric is analogous to the syndronous metric of standard cosmology, and ensures that all observers measure the same value of $l$, wherever they are located in the manifold. The latter includes the standard interval for spacetime given by $ds^2 = g_{\alpha\beta} dx^\alpha dx^\beta$, where there is summation of repeated up-and-down indices as usual. Below, we will sometimes refer to the hypersurface defined by 4D spacetime in the 5D manifold simply (if loosely) as $s$. The extra coordinate $x^4 = l$ is locally orthogonal to $s$. These considerations mean that we arrive at the 5D metric whose line element is given by

$$dS^2 = (l/L)^2 g_{\alpha\beta}(x^\gamma, l) dx^\alpha dx^\beta \pm dl^2.  \qquad (1)$$

We will see below that the lower sign here refers to classical physics, while the upper sign appears to refer to quantum physics

The equations of motion in 5D follow from extremizing the action via the formal relation $\delta\left[\int dS\right]=0$. However, we wish to make contact with known results, so we parametize the motion in terms of the 4D proper time $s$ and use the conventional normalization condition on the 4-velocities $u^\alpha \equiv dx^\alpha/ds$, namely $u^\alpha u_\alpha = 1$ or 0 for massive particles or photons. The canonical metric (1) then leads to a convenient split into equations for spacetime and the extra dimension. For a spacelike extra dimension, we find:

$$\frac{d^2 x^\mu}{ds^2} + \Gamma^\mu_{\alpha\beta} \frac{dx^\alpha}{ds}\frac{dx^\beta}{ds} = f^\mu \tag{2.1}$$

$$f^\mu \equiv \left(-g^{\mu\alpha} + \frac{1}{2}\frac{dx^\mu}{ds}\frac{dx^\alpha}{ds}\right)\frac{dl}{ds}\frac{dx^\beta}{ds}\frac{\partial g_{\alpha\beta}}{\partial l} \tag{2.2}$$

$$\frac{d^2 l}{ds^2} - \frac{2}{l}\left(\frac{dl}{ds}\right)^2 + \frac{l}{L^2} = -\frac{1}{2}\left[\frac{l^2}{L^2} - \left(\frac{dl}{ds}\right)^2\right]\frac{dx^\alpha}{ds}\frac{dx^\beta}{ds}\frac{\partial g_{\alpha\beta}}{\partial l}. \tag{2.3}$$

We see from (2.1) that the motion in spacetime is the standard geodesic one, modified by an extra force (per unit inertial rest mass) which is really an acceleration. It exists by (2.2) if the 4D metric depends on the extra coordinate and there is relative motion between the 4D and 5D frames. If either of $\partial g_{\alpha\beta}/\partial l$ or $dl/ds$ is zero, it vanishes. We have no reason at this stage to assume that there is zero motion in the extra dimension, so we recover conventional 4D mechanics if there is no intrusion of the fifth dimension into spacetime, whence $\partial g_{\alpha\beta}/\partial l = 0$ in (1), defining what is often called the pure-canonical metric. [Note that in this case, (2.3) has no solution with $dl/ds = 0$ except in the formal limit $l \to \infty$, so in general $dl/ds \neq 0$. It should also be noted that the normalization condition $g_{\alpha\beta}(x^\gamma, l) u^\alpha u^\beta = 1$ can be varied to show that the fifth force has the form (2.2) irrespective of the coordinate system in use.] The preceding comments mean that, dynamically speaking, the 4D Weak Equivalence Principle is the result of a 5D geometric symmetry.

To solve the equations (2) in full requires a solution of the field equations for $g_{\alpha\beta}(x^\gamma, l)$. However, we observe that the r.h.s. of (2.3) is zero irrespective of $\partial g_{\alpha\beta}/\partial l$ if

$$\frac{dl}{ds} = \pm \frac{l}{L} \quad . \tag{3}$$

The sign choice here merely reflects the reversibility of the motion and is not connected to the sign choice in (1). Now a solution of (2.3) in general involves two constants related to boundary conditions, which we take to be fiducial values of the parameters $l_*$, $s_*$ such that $l = l_*(s = s_*)$. We set $s_*$ to zero, whence the l.h.s. of (2.3) is zero if

$$l = l_* \exp(\pm s/L) \quad . \tag{4}$$

This is the required solution of (2.3). We see that the fifth force $f^\mu$ of (2.2) exists primarily because we are using $s$ and not $S$ to parametize the motion. [We could have $dl/ds \neq 0$ by (2.3) but $f^\mu = 0$ by virtue of $\partial g_{\alpha\beta}/\partial l = 0$ in (2.2), so we should not say that $dl/ds$ is caused by $f^\mu$.] Another interesting property of this analysis should be noted. So far, we have implicitly assumed $dS^2 \neq 0$ in the metric (1). But if we take (1) with the lower sign and set $dS^2 = 0$, we obtain $dl^2 = (l/L)^2 ds^2$ which again gives (3) and so (4). That is, these relations hold for a null geodesic, for which all particles in the 5D manifold are in causal contact. Yet another interesting property should be noted in conclusion. If we carry out the coordinate transformation $l \to L^2/l$, the metric (1) changes form, but if $dS^2 = 0$ then we obtain (3) and (4) yet again. In other words, there is a kind of duality between $l$ and $1/l$, to which we will later assign a physical meaning.

Some insight into the results of the previous analysis can be gained if we consider the Milne model of 4D FRW cosmology. The line element for this, in 3D isotropic coordinates, is given by

$$ds^2 = dt^2 - \left(\frac{t}{t_0}\right)^2 d\sigma^2, \quad d\sigma^2 \equiv \left[\frac{dr^2 + r^2 d\Omega^2}{\left(1 + kr^2/4\right)^2}\right], \quad (5)$$

where $t_0$ is a constant, $d\Omega^2 \equiv d\theta^2 + \sin^2\theta d\phi^2$ and the spatial curvature index is $k = -1$. Clearly this has the same form as (1) with a spacelike extra dimension, where the correspondences are between $t$ versus $l$ and $\sigma$ versus $s$. [In (5) a null path has $dt/t = \pm d\sigma/t_0$ while in (1) it has $dl/l = \pm ds/L$.] The essential properties of (5) are that it has negative 3D curvature and zero 4D curvature, so an observer confined to ordinary space would experience a non-Euclidean geometry, while one who could also move in time would find that the geometry was flat. [In fact, the Milne model can be derived from 4D Minkowski space by a well-known coordinate transformation, as given for example in ref. 28 p. 205, and accordingly it has $R_{\alpha\beta} = 0$ and is empty.] By analogy, the canonical metric (with spacelike $x^4$) has negative 4D curvature and zero 5D curvature, which is of course in con-formity with the field equations $R_{AB} = 0$. These equations have many cosmological solutions which are curved with matter in 4D but flat and empty in 5D [1, 16, 21, 22]. There is, though, one notable difference between the canonical metric and the Milne metric, in that the former generally implies a finite cosmological constant whereas the latter has none. To see this, we can take the pure-canonical metric $\left(\partial g_{\alpha\beta} / \partial l = 0\right)$ and reduce the 5D field equations $R_{AB} = 0$ to the 4D ones, which then take the form $G_{\alpha\beta} = 3g_{\alpha\beta} / L^2$ [1, 26]. This is the statement of an Einstein space with cosmological constant $\Lambda = 3/L^2$. This reduction also allows us to establish the Theorem: Any solution of the 4D Einstein equations without matter but with a cosmological constant can be locally embedded in a solution of the 5D Ricci equations with a (spacelike) pure-canonical metric. This applies especially to the Schwarzschild-deSitter 1-body solution and solutions for gravitational waves. (To be clear, here and below we mean by "matter" a fluid of conventional sort, with a density $\rho$ and pressure $p$ that are distinct from the equivalent description of the cosmological constant as a fluid with the equation of state $p = -\rho$, and possible scalar-field fluids of undetermined states.) The result just quoted is a particular case of Campbell's theorem [30], but is obviously of special physical importance.

Let us now consider the alternative case of the canonical metric (1), when it has a timelike extra dimension. The 5D interval, in terms of the 4D one and the extra coordinate, is now given by $dS^2 = (l/L)^2 ds^2 + dl^2$. Here $l$ plays the role of a second time, and it is known that such two-time metrics can have physical applications [10, 11]. One such application is to the basic theory of quantum mechanics, which employs an action

$$A \equiv \int mds = \int m\left[g_{\alpha\beta}\left(dx^\alpha/ds\right)\left(dx^\beta/ds\right)\right]ds = \int p_\alpha dx^\alpha \ . \tag{6}$$

Here $m$ is the (inertial) rest mass of a particle and $p_\alpha = mu_\alpha$ is its (covariant) momentum. It is conventional to use this to define a wave function, from which the momenta are derived in the usual manner:

$$\Psi \equiv \exp(-iA) \quad , \quad p_\alpha = (i/\Psi)(\partial\Psi/\partial x^\alpha) \quad , \tag{7}$$

where $p_\alpha p^\alpha = E^2 - p^2 = m^2$ for the energy and 3-momentum. We are recalling these elementary facts to make it clear that we can only hope to make contact with wave mechanics (and its more sophisticated developments) if we can introduce a complex quantity into classical relativity. And the most straightforward way to do this is to take a 5D metric of canonical type with signature $(+---+)$ and a null interval $dS^2 = 0$.

To see how this works, we can let $s \to is$ in our previous analysis for the $(+----)$ case. (The other way to change the signature is to let $l \to il$ and $L \to iL$, but this is cumbersome and runs into problems with the physical interpretation of these parameters given below.) Then the 4D geodesic given by (2.1) and (2.2) is still an equation in real quantities. As regards the motion in the extra dimension, (2.3) still has a zero r.h.s. because (3) now reads $dl/ds = \pm il/L$, and its l.h.s. is still zero because (4) now reads $l = l_* \exp(\pm is/L)$. That is, the dynamics remains valid, but now with a wave-like extra dimension. This causes the 4D curvature to become positive, and the effective cosmological constant to become negative and given by $\Lambda = -3/L^2$. (I.e., the space is of anti-deSitter type, as sometimes used in accounts of particle production by tunnelling in the early universe.) As regards the magnitude of $\Lambda$, we should recall that both here and above, it is a measure of the intrinsic curvature of spacetime; and this is in general distinct from its extrinsic curvature, which is measured by $l$ in the canonical metric. This is connected with the fact that in the literature some workers quote the magnitude of the 4D Ricci scalar as $12/L^2$ while some quote it as $12/l^2$, with corresponding magnitudes for $\Lambda$ of $3/L^2$ and $3/l^2$. (These quantities are related by $|R| = 4|\Lambda|$ of course: a similar ambiguity occurs in 4D scalar-tensor theory, where there is a choice between the so-called Einstein and Jordan frames.) Mathematically, the difference is almost trivial, since it just depends on the prefactor $l^2/L^2$ in the canonical metric (1). Physically, however, it may be significant because it depends on whether or not the observer can travel away from the spacetime surface $s$. An analogy is with the metric for 3D Euclidean space in spherical polars: $dr^2 + r^2(d\theta^2 + \sin^2\theta d\phi^2)$. For a 2D section, $r$ is often suppressed by putting it equal to unity, but this is not justified if one can travel away form the 2-surface. This question is important for the case of a timelike extra coordinate, because $l = l_* \exp(\pm is/L)$ necessarily involves excursions above and below $s$. We will return to this issue when we make a physical application of our results to quantization.

The shifted-canonical metric moves the events associated with (1) along the $l$-axis, by letting $l \rightarrow (l - l_0)$ where $l_0$ is a constant. This leaves the last part of (1) unaltered, but changes the prefactor on spacetime. This may appear to be a trivial shift, but it turns out to have significant consequences. This because the 5D and 4D coordinate transformations

$$x^A \to \bar{x}^A(x^B) \quad , \quad x^\alpha \to \bar{x}^\alpha(x^\beta) \tag{8}$$

are not equivalent. The first preserves $R_{AB} = 0$ while the second preserves $G_{\alpha\beta} = 8\pi T_{\alpha\beta}$. In general, 4D quantities (such as $T_{\alpha\beta}$) will change under a 5D coordinate transformation, making physics in spacetime gauge-dependent. To illustrate this, let us concentrate on the situation where the 4D metric is a conformally-flat function of the spacetime coordinates [12]. That is, we write $g_{\alpha\beta}(x^\gamma) = f(x^\gamma)\eta_{\alpha\beta}$ where $\eta_{\alpha\beta} = (+1, -1, -1, -1)$ is the metric of Minkowski space. This includes many cases of cosmological and astrophysical relevance, as noted elsewhere [28, 29]. With this understood, the 5D metric and its associated 4D cosmological constant are given by

$$dS^2 = \left(\frac{l-l_0}{L}\right)^2 g_{\alpha\beta}(x^\gamma)dx^\alpha dx^\beta - dl^2 \tag{9.1}$$

$$\Lambda = \frac{3}{L^2}\left(\frac{l}{l-l_0}\right)^2 . \tag{9.2}$$

We see that $\Lambda$ can now diverge, and only for $l \to \infty$ does it have its previous value $3/L^2$. In general, $\Lambda$ depends on the 4D proper time, because $l = l(s)$. This can be gotten either from the $l$-component of the geodesic (2.3), *or* directly for a null path from the metric (9.1). For the geodesic, we can use the solution $l = l_* \exp(\pm s/L)$ of (4) found above. Then (9.2) gives (for the upper sign) $\Lambda = (3/L^2)[1-\exp(-s/L)]^{-2}$, so the cosmological 'constant' decays from an unbounded value at the big bang ($s = 0$) to its standard value $3/L^2$ ($s \to \infty$). This and other aspects of the model are in agreement with astrophysical data [14]. The behaviour, however, is special in that $l(s)$ depends only on the arbitrary constant $l_*$ and not on the constant $l_0$ that measures the shift. A more general solution for $l = l(s)$ is obtained by using the null path $(dS^2 = 0)$ directly in the metric (9.1). There comes

$$l = l_0 + l_* \exp(\pm s/L) \quad . \tag{10}$$

This is for a spacelike extra dimension. A test particle clearly moves away from the hypersurface $l = l_0$ in the absence of non-gravitational forces (like electromagnetism), but at a slow rate governed by the cosmological constant $\Lambda = 3/L^2$. For a timelike extra dimension, the situation is different. Now we have

$$l = l_0 + l_* \exp(\pm is/L) \quad . \tag{11}$$

Clearly, a test particle now oscillates around the hypersurface $l = l_0$, with amplitude $l_*$ and wavelength $L$ (where now $\Lambda = -3/L^2$). That is, a change in the signature of the 5D metric causes a kind of confinement around 4D spacetime. We will use equations (10) and (11) below, in connection with classical and quantized dynamics.

3. Classical Mechanics and Cosmology

The forms of the canonical metric we examined in the preceding section can be used for models of real systems once we connect the (so far abstract) extra dimension to physics. This connection has proven controversial, hence the appearance over the years of several versions of 5D Kaluza-Klein theory. The first, by Klein in 1926, identified the momentum in the extra (compact) dimension with the quantum of electric charge $e$. But since this connection involved the quantum of action or spin momentum $\hbar$, it also led to particles of Planck mass ($10^{-5}$gm), which is contrary to observation. This is nowadays known as the hierarchy problem, and modern versions of Kaluza-Klein theory aim to avoid this, essentially by connecting the extra dimension not to the charge of a particle but to its mass *m*. This in membrane theory is proportional to the rate of change of the extra coordinate [7], while in space-time-matter theory it is proportional to the coordinate itself [6], though these approaches are equivalent by (3) if the metric is in canonical form. In what follows, we will be largely concerned with the role of mass in classical and quantum systems.

At a basic level, it is traditional to recognize three types of mass: the active gravitational mass $\left(M_g^a\right)$ which causes gravity, the passive gravitational mass $\left(M_g^p\right)$ which responds to gravity, and the inertial mass $\left(M_i\right)$ which measures the energy ($M_i c^2$: see refs. 1, 9, 27; in this section we replace the fundamental constants where necessary for physical understanding). The first two types are commonly identified by reciprocity arguments, and then the single gravitational mass is often set equal to the inertial mass by appeal to the Weak Equivalence Principle. However, a few astute workers have noted that what the WEP really implies is that $M_g$ and $M_i$ are *proportional* to each other. This will prove to be an important distinction, so let us consider briefly what is involved. For example, take the Earth (mass *m*) orbiting the Sun (mass *M*) and balance gravity with the centrifugal force as usual, with the appropriate type of mass. Then we obtain $GM_g m_g / r^2 = m_i v^2 / r$ for the velocity $v$ at radius $r$. This can be written in terms of dimensionless numbers as

$$\frac{GM_g}{c^2 r}\left(\frac{m_g}{m_i}\right) = \frac{v^2}{c^2} \quad . \tag{12}$$

A little thought about what is actually measurable shows that we can only assert that the factor in parentheses is a constant (not necessarily unity), or that gravitational mass and inertial mass are proportional to each other. This simple example has a counterpart in 5D relativity with an extra coordinate $x^4 = l$. We noted in Section 2 that the canonical metric for null paths implies a duality, insofar as $l \to L^2/l$ is a coordinate transformation that leaves the essential dynamics unchanged. Labelling the coordinates for these two gauges appropriately, we can in an algebraic fashion express this duality as $l_g l_i = L^2$. Here we recall that for the pure-canonical metric, the constant length here is related to the cosmological constant of general relativity via $\Lambda = 3/L^2$. There is a clear implication of these relations, which we need not insist on, but does serve the purpose of distinguishing between gravitational and inertial mass. For using the constants provided by nature, we can write $l_g = Gm_g/c^2$ and $l_i = h/m_i c$. That is, we can if we wish express the two types of mass via the Schwarzschild radius and the Compton wavelength. Then our duality reads $(Gm_g/c^2)(h/m_i c) = L^2$ or

$$\frac{Gh}{c^3}\left(\frac{m_g}{m_i}\right) = L^2 \quad . \tag{13}$$

This is analogous to (12) above, and like it, is a statement about the WEP. Note that if we were to insist on $m_g = m_i$, then by (13) we would have $L = (G\hbar/c^3)^{1/2}$, the Planck length, which is very small. But at least in a cosmological context, we have $L = (3/\Lambda)^{1/2}$, which is very large. This may be considered a restatement of the cosmological-'constant' problem [1, 13]. In the present approach, we instead take (13) to imply that we need to keep gravitational and inertial mass distinct as concepts, and that they can be represented by two different gauges in 5D relativity.

This inference is supported by calculating the energy of a soliton, which is a point source in empty, 3D spherically-symmetric 5D space. The soliton metric is a solution of the 5D Ricci-flat field equations $R_{AB} = 0$, and has been rediscovered by a number of workers in different coordinates [17]. It should be considered distinct from the Schwarzschild solution, which can be locally embedded in a pure-canonical metric by the theorem noted in Section 2 [1, 26]. Nevertheless, we can express the soliton metric in Schwarzschild-like coordinates, and for want of a better symbol denote the source by $M$ and define $A \equiv (1 - 2M/r)$. The field depends on two dimensionless constants $a$, $b$ which satisfy the consistency relation $(a^2 + b^2 + ab) = 1$, which is symmetric under the exchange of $a$ and $b$. These constants determine the soliton's energy $E_s$ which is in general not just $M$ [20]. Instead we have:

$$dS^2 = A^a dt^2 - A^{-a-b} dr^2 - A^{1-a-b} r^2 d\Omega^2 - A^b dl^2 \qquad (14.1)$$

$$E_s = M(a + b/2) \qquad . \qquad (14.2)$$

We see that the effective mass has two parts ($aM$ and $bM/2$), which are proportional to each other but in general not equal.

The canonical metric (1), as opposed to the soliton metric (14.1), is convenient because it can be explicitly related to the mass $m$ of a test particle as opposed to the mass $M$ of the source. Recalling that the form is $dS^2 = (l/L)^2 ds^2 - dl^2$, this can be done by expressing $m$ as a function of the extra coordinate $x^4 = l$. Indeed in its standard form, the canonical metric implies the identification $m = l$ for several reasons: (a) the first part of the 5D metric then gives the element of classical action $mds$; (b) the constant of the motion for the time axis then has its usual form $m(1 - v^2/c^2)^{-1/2}$ where $v$ is the 3D velocity. In addition, one may note for any form of the metric that: (c) if $l$ does not appear, the effective 4D energy-momentum tensor has zero trace, signifying particles with zero rest mass, as expected; (d) all of mechanics involves the three physical base quantities $M, L, T$ used in dimensional analysis, so a complete metric-based theory ought to include a mass dimension as well as those for length and time. These four comments do not exhaustively prove the correspondence between $m$ and $l$, but do strongly suggest it.

If the mass of an object can be considered as a coordinate (either via $m = l$ or $m = 1/l$ depending on the gauge), it is natural to inquire about the laws of dynamics. This subject has been considered at length elsewhere [1, 4-8, 26]. Here, we recall from Section 2 that for both membrane theory and induced-matter theory, there is in general a fifth force (per unit mass) due to the fifth dimension. For the canonical metric, it is given by (2.2). Since it acts parallel to the 4-velocity $u^\alpha \equiv dx^\alpha/ds$ and is proportional to $dl/ds$, we would in 4D interpret this as an anomalous change in the rest mass $m$. However, in practice this does not usually upset conventional dynamics. For in the pure-canonical case $\left(\partial g_{\alpha\beta}/\partial l = 0\right)$ as required by the WEP, the fifth force is absent, leaving the accelerations exactly as they are in conventional 4D theory. (This applies to the canonical embedding of the Schwarzschild solution, and so to the solar system and other astrophysical situations.) Also, in most other cases $\left(\partial g_{\alpha\beta}/\partial l \neq 0\right)$, the fifth force causes the mass and the velocity to change in inverse proportion to each other, so preserving the momentum. Further investigation shows that there *are* certain circumstances where the fifth force can modify the standard laws of dynamics, but we defer commenting on this until after we discuss quantization.

Here, we note that while classical systems usually have conventional geodesic motion in 4D, they can still process properties which are special to 5D. The main such is the *l*-behaviour given by (4), which implies a variation of rest mass. However, this is very slow, since $(1/l)(dl/ds) = \pm 1/L = \pm(\Lambda/3)^{1/2}$, and has a cosmological timescale. Also, as noted above, such a variation cannot be detected dynamically if the metric is pure-canonical ($\partial g_{\alpha\beta}/\partial l = 0$ after *l*-factorization). As regards the solar system – which historically has provided the main testbed for gravitation – it should be appreciated that what we verify when we compare theory and observation is essentially an *orbit*. In the case of a planet going around the Sun, this is close to an ellipse. The orbit in plane polar coordinates is given by $r = a(1-e^2)(1+e\cos\theta)^{-1}$, where $a$ is the semimajor axis and $e$ is the eccentricity. In other words, what we are testing is a relation between coordinates, in this case $r = r(\theta)$. We mention this, because in more complicated cases in 5D non-compactified theory, the orbit may not be confined to the 4D hypersurface usually called spacetime.

Classical universe models have been much studied. The traditional approach to this subject has been to look for solutions of the field equations $(R_{AB} = 0)$ in 5D which have metrics of Friedmann-Robertson-Walker type in 4D. In this way, 5D analogs were found for the vacuum-dominated, radiation-dominated and matter-dominated phases of standard cosmology [1, 21, 22]. Much of this work was done before the introduction of canonical coordinates. However, we should recall that in principle any 4D system can be described by the general canonical metric (1); and that any system close to vacuum can be approximated by its pure form ($\partial g_{\alpha\beta} / \partial l = 0$). One important solution with a pure-canonical metric is the 5D analog of the Milne universe (5), but with a cosmological constant added. It is an exact solution of $R_{AB} = 0$, and has 5D line element

$$dS^2 = \left(\frac{l}{L}\right)^2 \left\{ dt^2 - \left[L \sinh\left(\frac{t}{L}\right)\right]^2 d\sigma^2 \right\} - dl^2 \quad, \tag{15}$$

where the 3-space has $d\sigma^2 = (dr^2 + r^2 d\Omega^2)(1 + kr^2/4)^{-2}$ with $k = -1$. This has no ordinary matter, but a cosmological constant $\Lambda = 3/L^2$. It is a viable model for the very early inflationary universe, as is its $k = 0$ counterpart which is a 5D embedding of the 4D deSitter solution. Both have a vacuum with the classical equation of state $p = -\rho$, so the gravitational mass density $(\rho + 3p)$ is negative and powers a strong expansion.

Later, the universe is believed to have gone through a period when its equation of state was close to that of radiation $(p = \rho/3)$, before evolving into a state similar to dust ($p = 0$). The same 5D metric can describe both of these phases, and is commonly taken in the (non-canonical) form where it reduces to the appropriate 4D Robertson-Walker one on hypersurfaces $x^4 = l$ [21]. The line element is

$$dS^2 = l^2 dt^2 - t^{2/\alpha} l^{2/(1-\alpha)} \left( dr^2 + r^2 d\Omega^2 \right) - \alpha^2 (1-\alpha)^{-2} t^2 dl^2 \quad . \tag{16.1}$$

This has ordinary matter, where in terms of the cosmic time $\tau$ the density and pressure are given by

$$8\pi\rho = \frac{3}{\alpha^2 \tau^2} \,, \qquad 8\pi p = \frac{2\alpha - 3}{\alpha^2 \tau^2} \quad . \tag{16.2}$$

The equation of state is the isothermal one, $p = (2\alpha/3 - 1)\rho$. For $\alpha = 2$, the scale factor of (16.1) varies as $t^{1/2}$, (16.2) gives $\rho = 3/32\pi\tau^2 = 3p$ and we have the standard radiation model for the early universe. For $\alpha = 3/2$, the scale factor varies as $t^{2/3}$, the density and pressure are $\rho = 1/6\pi\tau^2$ with $p = 0$, and we have the standard dust model for the late universe.

A remarkable discovery was made about the last-quoted metric (16.1) some time after its formulation: it is not only Ricci-flat $(R_{AB} = 0)$ but also Riemann-flat $(R_{ABCD} = 0)$. In other words, there is a set of coordinates in which (16.1) becomes the metric of flat Minkowski space in 5D:

$$dS^2 = dT^2 - R^2 \left( d\theta^2 + \sin^2\theta d\phi^2 \right) - dL^2 \quad . \tag{17}$$

The precise transformations between *T, R, L* and *t, r, l* are complicated but known (ref. 1 p.37). They allow of the curved 4D universe to be plotted in flat 5D space, aiding visualization. It should also be mentioned that the Milne-like metric (15) for the very early universe is also 5D flat. These discoveries let to several studies of cosmological and astrophysical metrics which are flat in 5D but curved in 4D [23]. The related embeddings are governed by Campbell's theorem [30], and the 4D matter is given by the induced energy-momentum tensor of space-time-matter theory [1, 18, 22]. Of course, not all solutions of $R_{AB} = 0$ also have $R_{ABCD} = 0$, and the solitons with metric (14) are such. It is the high degree of symmetry of the 4D RW metric which enables it to be embedded in 5D Minkowski space, and it is now known that all FRW models can be so treated. This means that the universe which is curved and contains matter in 4D may – if we wish – be regarded as flat and empty in 5D.

4. <u>Quantum Mechanics and Particles</u>

Quantized models for particles would appear at first sight to involve a different breed of physics from what we just discussed. In Section 3, we noted that there are several versions of the canonical metric (1). These include the pure case ($\partial g_{\alpha\beta}/\partial l = 0$), the transformed case ($l \to L^2/l$) and the shifted case ($l \to l - l_0$). Also, we commented that the sign of the last term in the metric can be negative or positive, corresponding to a spacelike or timelike extra coordinate respectively (with the cosmological constant being positive or negative respectively). These choices of coordinate frame – or gauge – are all allowed, because the underlying theory is covariant in 5D. However, because of the dif-ference in the 5D and 4D groups of coordinate transformations (8), we expect that if quantization shows up at all, then it may only do so in some and not all of the noted cases. In other words, since the standard rules of quantization are 4D in nature, then the concept is itself gauge-dependent. This proves to be so. In the present section, we therefore concentrate on two cases of the canonical metric where quantization occurs. Since our results are surprising and preliminary, we will proceed succinctly, just stating the relevant metrics and outlining how they lead to quantization. Detailed analyses can be deferred, and for now we note the following two cases:

(A) For a spacelike extra dimension, we can transform the canonical metric via $l \to L^2/l$ and choose to measure the rest mass of a test particle in the ($x^\alpha, l$) manifold by $l = h/mc$. Then for a 5D null-path we obtain:

$$0 = dS^2 = (L/l)^2 ds^2 - (L/l)^4 dl^2 \qquad (18.1)$$

$$d(L/l) = \pm ds/l \quad . \qquad (18.2)$$

The last relation here comes from (18.1) and is an alternative form of (3). The sign choice merely reflects the reversibility of the motion in the fifth dimension, as noted previously. Let us suppress this sign choice, define $n \equiv L/l$ and restore physical units. Then (18.3) says

$$\int mc\, ds = nh \quad , \qquad (18.3)$$

which is the usual rule for quantization when $n$ is an integer (this in the sense of the older version of Bohr-Sommerfeld quantization). We see that quantization occurs if the extra coordinate $x^4 = l$ 'fits' integrally into the intrinsic lengthscale $L$ of the geometry. This implies structure in the fifth dimension, analogous to the compactification of the original Klein model [1]. In the latter, the electron charge $e$ was taken to be proportional to the velocity in the extra dimension $dl/ds$. In the present approach, we are focusing on rest mass rather than electric charge, but (18.3) implies the proportionalities $e \sim dl/ds \sim l/L \sim 1/n$. That is, charge is also quantized, but large objects are neutral in the classical limit ($n \to \infty$). The model we are discussing in the form of equations (18) is physically acceptable, but leaves unanswered the question of why the structure implied by $L/l = n$ should exist.

(B) For a timelike extra dimension, we can transform the metric as before and choose to measure the mass by $l = \hbar/mc$. The use of $\hbar \equiv h/2\pi$ here reflects the fact that the 4D curvature of spacetime changes sign by virtue of the change in the signature of the metric, admitting closed 4D spaces around which a 4D wave may run with angular period $2\pi$ in the $l/s$ plane. (The 4D Ricci scalar as defined in terms of its embedding in 5D is given e.g. on p. 16 of ref. 1, and changes sign when the fifth dimension changes from spacelike to timelike.) The difference is trivial mathematically, but is significant physically because the compact nature of the spacetime surface $s$ gives a rationale for the 5D structure of the present model that was lacking in its predecessor (A). Then for a 5D null-path we obtain

$$0 = dS^2 = (L/l)^2 ds^2 + (L/l)^4 dl^2 \qquad (19.1)$$

$$m = m_* e^{\pm is/L} \qquad (19.2)$$

The last relation here comes from (19.1) and is an alternative (complex) form of (4). It says that the mass of a test particle in the 5D manifold is wavelike, oscillating with wave-length $L$ and amplitude $m_*$ around zero. If we wish to shift the locus of the motion to a finite value, we can as before do $l \to (l - l_0)$ so that $l = l_0 + l_* \exp(\pm is/L)$, a situation illustrated in Fig. 1. Taking the orbit to be closed in the $l/s$ plane, where the element of phase is $d\theta = ds/l$, we can consider $n$ orbits of $2\pi$ radians each. Then we obtain

$$2\pi n = \int_0^{2\pi n} d\theta = \int \frac{ds}{l} = \int \frac{mc\,ds}{\hbar} \quad , \tag{19.3}$$

which again gives the usual rule for quantization or $\int mc\,ds = nh$ in terms of the straight value of Planck's constant.

In the preceding models (A) and (B), we have sketched how conventional 4D quantization might arise from structure in 5D. It is important to realize, however, that we have only described two cases out of many possible gauge choices. But that said, a timelike extra dimension appears favoured, because it leads automatically to confinement; and makes the mass wavelike, so providing a rationale for wave-particle duality. That is, a timelike extra dimension allows the 4-velocities $u^\alpha \equiv dx^\alpha / ds$ to be converted via a wavelike $m$ to deBroglie waves for the energy and 3-momentum. This solves a long-standing conundrum in mechanics. A detailed investigation of wave-particle duality should be carried out in the context of 5D relativity with a mass-related extra coordinate. Certain technical problems will have to be considered in such an analysis. For example, conventional causality as defined by $ds^2 \geq 0$ should presumably be maintained. Also, $ds^2$ is a quadratic form, whereas the representation of a complex quantity in terms of a real part and a wave part only strictly applies to a linear form, so there may be a problem with su-perposition. Lastly, the charge/parity/time invariance of 4D theory will have to be extended to 5D by including the mass in a CPTM theorem.

Intriguingly, some aspects of the preceding account also appear in an analysis of 1935 by Dirac [32]. He gave a neat classification of the energy and 3-momenta of a particle by embedding 4D deSitter space in a 5D manifold. He was led to the conclusion that the rest mass of a test particle may be a complex quantity. His work was formal in nature, and only touched on the cosmological 'constant'. By contrast, the energy density and pressure of the vacuum (as measured by $\Lambda$) lie at the heart of the present account. For if mass can be regarded as a wave, it is necessarily a wave in the vacuum.

In 5D relativity, the magnitude of $\Lambda$ is determined by the inverse square of a length associated with the metric [1, 8, 26, 31]. This length is $L$ as measured intrinsic to a 4D hypersurface (i.e. by Einstein's equations), or $l$ as measured extrinsic to it (i.e. by the mass in the canonical metric). The mismatch in these two ways of measuring $\Lambda$ offers a way to reconcile the disparate measures of this parameter from cosmology and particle physics [1, 13]. We note that the lengthscale of the observable universe is of order $10^{28}$cm, while the Compton wavelength of the proton is of order $10^{-12}$ cm. Thus

$$\frac{\Lambda(\text{proton})}{\Lambda(\text{universe})} \approx \left(\frac{L}{l}\right)^2 = \left(\frac{10^{28}}{10^{-12}}\right)^2 = 10^{80} \quad . \tag{20}$$

This is in the range $10^{60} - 10^{120}$ usually quoted for the discrepancy in the value of the cosmological 'constant'. It is actually the same order of magnitude as the number of protons in the observable universe, a coincidence which invites further study.

5. Summary and Discussion

We have taken the simplest extension of general relativity from 4 to 5 dimensions, and shown that a spacelike and timelike extra dimension lead respectively to models of classical and quantum systems. In the former domain, much is already known, including 1-body and cosmological models that are in agreement with observation. In the latter domain, we have seen that a change in the signature of the metric is necessary in order to recover the usual description in terms of the complex quantities of particle physics. The two domains are also connected by a coordinate transformation or change of gauge when the canonical metric is used. This duality can be related to the logical distinction between gravitational and inertial mass when the extra coordinate is related to rest mass. (For non-canonical metrics, the fifth dimension is still connected to mass, but not in so direct a manner.) The Weak Equivalence Principle in a dynamical sense is obeyed when the spacetime segment of the metric is independent of the extra coordinate and /or the 4D frame does not move with respect to the 5D one. The cosmological-'constant' problem can be avoided, essentially because this parameter in 5D can be variable, being larger in magnitude for particles smaller in mass. We have concentrated on dynamics rather than field equations, though many exact solutions are known of these in the Ricci-flat form $R_{AB} = 0 \, (A, B = 0-4)$. A natural assumption from these field equations is that all particles in 4D, irrespective of their masses, follow null paths in 5D. (That is, $dS^2 = 0$ for the 5D interval.) In this theory, photons are those particles which are confined to the massless 5D hypersurface we call spacetime. Massive particles can oscillate around the hypersurface, in a way which combines a wavelike mass with the 4-velocities to explain deBroglie waves and wave-particle duality. The physics of massive particles flows largely from the null-path hypothesis, which is the main assumption in the above analysis.

This condition implies that all objects in the 5D universe are in some sense in causal contact with each other (analogous to how $ds^2 = 0$ defines communication in 4D). This is in agreement with the fact that the particle properties of remote objects such as quasars are measured to be the same, even though such sources were out of causal contact with each other, early on in certain 4D models of cosmology [24]. The accuracy of this uniformity needs to be studied in detail, from both the observational and theoretical perspectives. More work is also required on the connection between the cosmological 'constant', as a measure of the classical vacuum field in 4D, and the scalar field which exists in 5D [1, 13]. In the canonical metric which we have mainly used in the above, this field is taken to be uniform in space and time. Relaxing this condition will lead among other things to a modification of the laws of motion in a classical sense. It may also have important implications for particle physics. The 5D theory in a quantum-field sense involves a spin-2 graviton, a spin-1 photon and a spin-0 scalaron. In regard to the last, it is sometimes stated that the extra potential in Kaluza-Klein theory plays the role of the Higgs field in quantum theory, and that the magnitude of the Higgs field is related to the size of the cosmological constant. The precise relationship between these subjects needs to be elucidated from the theoretical side. The nature of the 5D scalar field, as well as the strength of the fifth force, can also be investigated from the experimental side, notably in the planned satellite test of the Equivalence Principle.

Given the scope of work yet to be done, the present account should be termed preliminary in nature. However, it is remarkable that so much classical and quantum physics has been deduced from the swapping of a sign in the metric and the idea that particles travel on null paths in a higher-dimensional space.

Acknowledgements

This paper is based on previous work with B. Mashhoon, H Liu and other members of the Space-Time-Matter consortium (webpage http://astro.uwaterloo.ca/~wesson). It was partly supported by N.S.E.R.C.

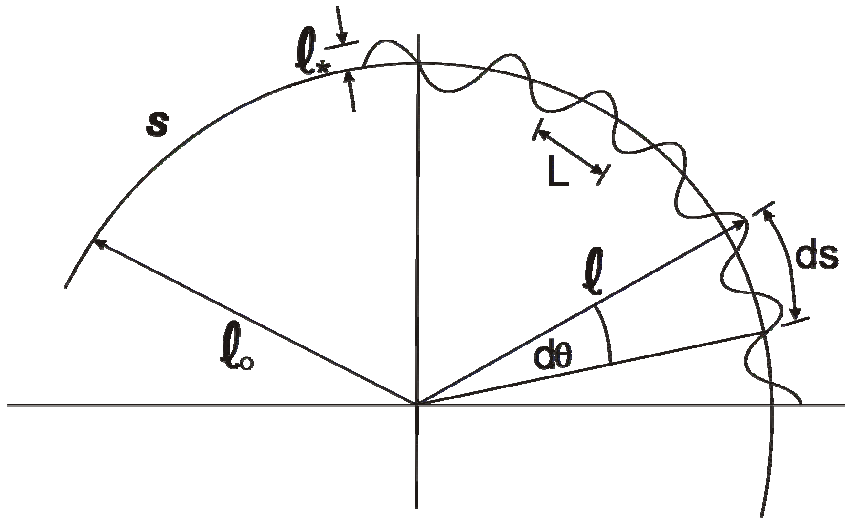

Figure Caption

To illustrate the propagation of a wave in the fifth coordinate $(l)$ around space-time $(s)$. The wave has an amplitude $l_*$ and a wavelength $L$, and an increment in the phase is determined by the 4D interval via $d\theta = ds/l$. The mean size of the orbit is $l_0$, and while $s$ is closed its cross-section need not be circular. Not to scale.